\documentclass[sigconf,nonacm]{acmart}

\usepackage{titlesec}

\titleformat{\subsubsection}
  {\normalfont\normalsize\bfseries}
  {\thesubsubsection}
  {1em}
  {}

\titlespacing*{\subsubsection}
  {0pt}
  {1.5ex plus .2ex}
  {1ex plus .2ex}

\microtypesetup{expansion=false}

\renewcommand\footnotetextcopyrightpermission[1]{}
\usepackage{amsmath}
\usepackage{tabularx}
\usepackage{booktabs}
\usepackage{graphicx}
\usepackage{float}
\usepackage{listings}
\usepackage{xcolor}

\title{LLM-Guided Reinforcement Learning for Adaptive NPC Behavior in Multi-Agent Combat Games}

\author{Hrithika Deepu Nair}
\affiliation{%
  \institution{Heriot-Watt University}
  \city{Dubai}
  \country{U.A.E}
}
\email{hrithikanair2003@gmail.com}

\author{Kayvan Karim}
\affiliation{%
  \institution{Heriot-Watt University}
  \city{Dubai}
  \country{U.A.E}
}
\email{k.karim@hw.ac.uk}

\begin{document}

\begin{teaserfigure}
    \centering
    \includegraphics[width=0.85\textwidth]{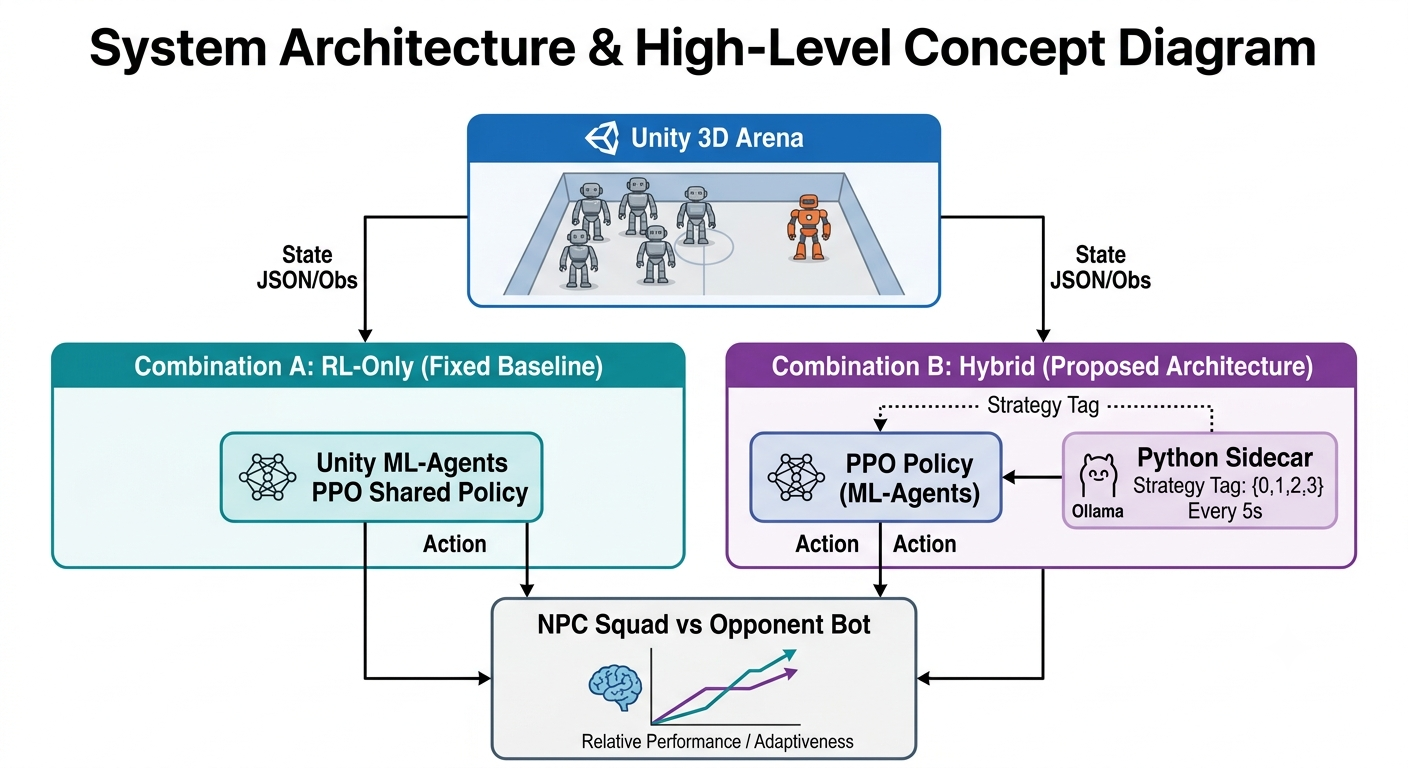}
    \caption{System overview comparing Combination A (RL-only baseline) and Combination B (RL plus Mistral 7B strategy selector via Ollama), both evaluated in the same Unity 3D arena.}
    \label{fig:intro_overview}
\end{teaserfigure}
 
\begin{abstract}
Non-player characters in combat video games have long frustrated players and designers alike. Scripted enemies follow fixed patterns that experienced players learn to exploit within minutes, whilst purely rule-based systems offer no mechanism for adjusting to what a specific opponent is actually doing. Reinforcement learning offered a partial solution, producing agents that learn effective behaviour through millions of simulated encounters, yet once training ends the policy is frozen. The agent that performed well against one opponent type applies the same strategy against every other, with no way to recognise the difference.
 
Large language models changed what seemed possible. Systems capable of reading a situation and reasoning about the best course of action without task-specific training opened a new design space: what if strategy selection were handed to a language model, leaving the trained policy to handle only execution? Prior work explored language models as reward designers during training and as players in turn-based text games, but no study had tested whether a locally hosted language model querying a live game state in real time could make reinforcement-learning agents measurably more adaptive during actual play.
 
This paper tests that question directly. Five NPC agents share a single PPO policy trained in Unity. In the baseline configuration the policy acts alone. In the experimental configuration, a Python sidecar runs a locally hosted Mistral 7B model via Ollama, reads the live game state every five seconds, and assigns one of four tactical tags that conditions how the shared policy behaves. Three scripted opponent bots enable controlled measurement across 600 evaluation episodes, compared statistically using the Mann-Whitney U test.
 
The results are condition-dependent. Against a Balanced opponent that switches tactics mid-episode, the augmented squad more than doubled its win rate from 11 to 24 per cent and produced significantly longer episodes. Against an Evasive opponent it won more often and killed faster, though shorter episode duration places this outside the strict hypothesis definition. Against an Aggressive opponent the language model's near-constant preference for encirclement proved counterproductive. Analysis of 2,430 tag selections revealed an 83.8 per cent Surround preference regardless of opponent type, indicating limited zero-shot strategic differentiation at this model scale. The finding characterises both the promise and the current ceiling of language-model-guided runtime strategy selection as a design pattern for adaptive game AI.
\end{abstract}

 
 
\maketitle
 
\section{Introduction}
\label{sec:intro}
 
The central question is whether a language model reading a live game state every five seconds can make a squad of trained agents meaningfully more adaptive than the same squad operating without one.
 
\subsection{Background and Motivation}
\label{sec:intro_motiv}
 
Artificial intelligence (AI) in video games has long relied on scripted behaviour trees and finite state machines (FSMs) to govern non-player character (NPC) actions. While these approaches are computationally efficient and deterministic, they produce NPCs whose patterns experienced players quickly learn to exploit~\citep{bontchev2024}. As player expectations have grown, the demand for NPCs that feel genuinely adaptive, capable of responding to player strategies, switching tactics mid-battle, and coordinating with teammates has become a central challenge in game AI research.
 
Reinforcement learning (RL) has emerged as a powerful alternative, enabling agents to learn effective combat behaviour through thousands of simulated episodes of trial and error. Recent breakthroughs in deep RL, such as the AlphaStar system \citep{vinyals2019} and OpenAI Five \citep{berner2019}, demonstrate that RL-trained agents can achieve superhuman performance in complex strategy games. However, these systems require enormous computational resources and, critically, learn a fixed policy that does not easily adapt at runtime to novel opponent behaviours without additional retraining.
 
Large Language Models (LLMs) provide a complementary form of capability. They are trained on extensive datasets that capture human knowledge and patterns of reasoning. LLMs can process natural language descriptions of game states, reason about high-level strategic options, and output structured decisions without task-specific fine-tuning. Recent work has begun to explore LLMs as reward designers \citep{li2024auto}, game-playing agents \citep{akata2025playing}, and strategic advisors for RL systems \citep{afonso2025self}. This research sits at the intersection of these threads, asking whether an LLM acting as a real-time strategy selector can meaningfully increase the adaptiveness of RL-trained combat NPCs in a squad-based battle scenario.
 
To make this concrete, consider the following illustrative scenario. At the 30-second mark of a battle between the five-NPC squad (Combination B) and an evasive opponent bot, the game state reads: bot HP 78\%, average NPC HP 52\%, and three NPCs within 6 units of the bot. The LLM sidecar receives this as a structured JSON prompt, reasons that the bot is healthy, mobile, and avoiding engagement, and returns tag 0 (Surround). All five NPCs update their conditioning input simultaneously, and the shared PPO policy, which has been trained to maximise angular spread around the target, repositions agents to encircle the bot. Ten seconds later, with bot HP at 41\% and retreating toward a wall, the LLM switches to Tag 1 (Aggressive), triggering the policy to close in and press the advantage. In Combination A (no LLM), the NPCs apply the same fixed average behaviour throughout, with no mechanism for this kind of mid-battle strategic pivot. This contrast defines the empirical question at the heart of this research.
 
\subsection{Research Questions and Hypothesis}
\label{sec:intro_hypo}
 
This research is structured around the following primary research question:
 
\begin{quote}
\textbf{RQ:} To what extent does LLM-guided real-time strategy selection produce measurably greater NPC adaptiveness --- operationalised as time-to-defeat --- compared to a shared PPO baseline in a controlled multi-agent combat environment?
\end{quote}
 
Two hypotheses are tested:
 
\begin{itemize}
\item \textbf{H1:} Combination B (RL + LLM strategy selector) will produce statistically significantly longer mean time-to-defeat than Combination A (RL only), across all three opponent bot types.
\item \textbf{H0:} There is no statistically significant difference in mean time-to-defeat between Combination A and Combination B.
\end{itemize}
 
Statistical comparison is performed using the two-sided Mann-Whitney U test with effect size reported as rank-biserial correlation $r$. The Mann-Whitney U test is chosen because it is non-parametric and makes no assumption of normality in the time-to-defeat distributions, which are bounded between zero and the 120-second episode limit and are unlikely to be normally distributed given the discrete nature of episode outcomes~\citep{conover1999}. H0 is rejected when $p < 0.05$.
 
\subsection{Aim and Objectives}
\label{sec:intro_aim}
 
The overarching aim of this project is to empirically evaluate whether integrating an LLM as a high-level strategy selector improves the adaptiveness of RL-trained NPC agents in a multi-agent combat environment, as measured by time-to-defeat. The specific objectives are:
 
\begin{enumerate}
\item[\textbf{O1.}] Design and implement a Unity-based combat environment featuring five NPC agents trained under PPO using the ML-Agents toolkit.
\item[\textbf{O2.}] Establish a baseline RL-only configuration (Combination A) using a shared policy trained without LLM guidance, with step-based automatic bot-type rotation during training.
\item[\textbf{O3.}] Implement an LLM-augmented configuration (Combination B) in which a Mistral 7B sidecar assigns high-level strategy tags every five seconds that modulate the shared PPO policy.
\item[\textbf{O4.}] Develop three scripted opponent bots (Aggressive, Evasive, Balanced) to enable repeatable, controlled evaluation.
\item[\textbf{O5.}] Statistically compare the two configurations across 300 evaluation episodes per combination using the Mann-Whitney U test.
\item[\textbf{O6.}] Critically evaluate whether the LLM-augmented system demonstrates statistically significant improvement in adaptiveness and examine the conditions under which LLM guidance does or does not add measurable value.
\end{enumerate}
 
\subsection{Contributions}
\label{sec:contributions}
 
This paper makes the following contributions to the field of game AI and LLM-RL integration.
 
\begin{enumerate}
\item[\textbf{C1.}] \textbf{Controlled empirical evaluation of runtime LLM-guided strategy selection.} This work isolates runtime LLM guidance as the only experimental variable by comparing an RL-only shared PPO policy with an otherwise identical LLM-augmented architecture, enabling the effect of online strategy selection on NPC adaptiveness to be measured directly.
 
\item[\textbf{C2.}] \textbf{Empirical characterisation of runtime decision behaviour in a local 7B language model.} Analysis of 2,430 Mistral 7B strategy selections reveals a dominant \textit{Surround} preference (83.8\%) across all opponent types, providing quantitative evidence of limited zero-shot strategic differentiation.
 
\item[\textbf{C3.}] \textbf{Identification of opponent-dependent effects of runtime LLM guidance.} The evaluation shows that runtime LLM strategy selection improves performance against some opponent types but reduces it against others, demonstrating that its effectiveness depends on the combat scenario rather than being universally beneficial.
 
\item[\textbf{C4.}] \textbf{Analysis of reward-shaping interactions in strategy-conditioned shared policies.} A reward-shaping failure mode during training is identified, along with a practical modification that restores stable learning while preserving the experimental design.
\end{enumerate}
 
\subsection{Organisation}
\label{sec:intro_orga}
 
Section~\ref{sec:intro} introduces the project; states the research question and hypotheses alongside the statistical test used to evaluate them and the contributions; and defines the aims and objectives. Section~\ref{sec:literature} reviews the relevant literature across five areas: background and technical foundations, RL in games, multi-agent RL, LLM integration with RL, and NPC behavior adaptation, culminating in a structured comparison table and a precise statement of the research gap. Section~\ref{sec:method} presents the full research design and implementation, covering the game environment, animation system, project requirements, policy architecture, reward function, LLM sidecar, opponent bot designs, and evaluation protocol. Section~\ref{sec:results} presents evaluation results, descriptive statistics, and statistical analysis. Section~\ref{sec:discussion} and Section~\ref{sec:conclusion} draw conclusions, discuss limitations, and identify future work.
 
\section{Related Work}
\label{sec:literature}
 
None of the design decisions in Section~\ref{sec:method} were made in a vacuum, and the reasoning behind each is only convincing once the underlying literature has been examined critically rather than cited in passing. Each area reviewed below corresponds to a point where this project had to choose between competing approaches; each design choice is evaluated against the alternatives the existing literature offers.
 
\subsection{Background and Technical Foundations}
\label{sec:literature_back}
 
The design choices defended later in this paper lean on a small set of concepts --- MDPs, policy gradients, transformers, and reward shaping --- that are worth stating precisely once here, rather than assuming or redefining them loosely each time they recur.
 
\subsubsection{Reinforcement Learning}
\label{sec:literature_back_rl}
 
\textbf{Markov Decision Process.} Reinforcement learning (RL) is a branch of machine learning in which an agent learns to take actions in an environment by receiving feedback in the form of scalar reward signals. Formally, the problem is modelled as a Markov Decision Process (MDP) defined by the tuple $(S, A, P, R, \gamma)$, where $S$ is the state space, $A$ is the action space, $P$ is the transition probability function, $R$ is the reward function, and $\gamma \in [0,1]$ is the discount factor \citep{sutton1998}. The agent's goal is to learn a policy $\pi : S \rightarrow A$ that maximises the expected cumulative discounted reward:
\begin{equation}
  G_t = \sum_{k=0}^{\infty} \gamma^k r_{t+k+1}
  \label{eq:rl_return}
\end{equation}
Game environments satisfy the Markov property only approximately, since agents with partial observations cannot see the full joint state. This project addresses that by including ally positions and HP in each NPC's observation vector, giving each agent a team-level approximation that a single-agent formulation would not provide.
 
\textbf{Proximal Policy Optimization.} Proximal Policy Optimization (PPO) \citep{schulman2017} is a policy gradient algorithm that directly optimises $\pi$ by performing gradient ascent on an objective clipped to prevent excessively large policy updates. It maintains a ratio:
\begin{equation}
  r_t(\theta) = \frac{\pi_{\theta}(a_t \mid s_t)}{\pi_{\theta_{\text{old}}}(a_t \mid s_t)}
  \label{eq:ppo_ratio}
\end{equation}
and clips this ratio to $[1-\epsilon, 1+\epsilon]$ with $\epsilon=0.2$. PPO is the default algorithm in Unity ML-Agents~\citep{juliani2018} and is used throughout this project, chosen over SAC or DQN for its native support, established convergence behaviour in discrete action spaces, and on-policy nature, which suits a stationary scripted opponent where off-policy experience offers little advantage.
 
\subsubsection{Large Language Models}
\label{sec:literature_llm}
 
Large Language Models (LLMs) are neural networks trained via self-supervised next-token prediction on large text corpora, built on the transformer architecture \citep{vaswani2017attention}, which uses multi-head self-attention to model token dependencies across arbitrary sequence lengths. Pre-training on diverse text gives LLMs emergent capabilities, including instruction following and zero-shot generalisation, performing tasks with no task-specific examples, guided only by a natural-language description. This is what makes LLMs viable as runtime strategy selectors here without game-specific fine-tuning.
 
This is an empirical property, however, not a guaranteed one, and whether a locally hosted 7B model retains it for spatial strategic reasoning specifically is not established by the literature reviewed here. This uncertainty is revisited as a limitation in Section~\ref{sec:critical_eval}.
 
\subsubsection{Reward Shaping}
\label{sec:literature_back_reward}
 
Reward shaping augments the environment's natural reward signal to accelerate or guide RL training. A shaping function $F(s, a, s') = \gamma \Phi(s') - \Phi(s)$ is added to the base reward to encourage desired behaviours without changing the optimal policy under mild conditions \citep{ng1999potential}, a guarantee later extended to shaping functions that vary dynamically between episodes \citep{devlin2012} directly relevant here, since Combination B samples a different strategy tag and, therefore a different shaping function, at the start of each training episode. Reward shaping in this project serves a purpose beyond accelerating training: it is what teaches the shared policy what each strategy tag means spatially, since the base reward function alone gives no incentive to behave differently under different tags. The full shaping formulation is detailed in Section~\ref{sec:reward_function}.
 
\subsection{Reinforcement Learning in Game Environments}
\label{sec:literature_rl}
 
Theory alone does not justify choosing RL for this project; what matters is whether it has actually worked in comparable settings, and at what cost. The systems reviewed below span several orders of magnitude in scale, and the pattern that recurs across all of them --- capability without runtime flexibility --- is what makes the rest of this paper necessary.
 
\subsubsection{Deep RL Landmark Systems}
 
Deep RL became viable for sequential decision-making once a Deep Q-Network (DQN) was shown to learn 49 Atari games directly from pixel observations \citep{mnih2015}. Large-scale systems extended this dramatically: OpenAI Five \citep{berner2019} and AlphaStar \citep{vinyals2019} reached expert-level performance in Dota 2 and StarCraft II through extensive self-play. Both represent the upper bound of what deep RL can achieve but share a limitation central to this project's motivation: once trained, they execute a fixed policy. In Dota 2, the high-level strategy was hard-coded, with no mechanism for switching strategy mid-game without retraining \citep{berner2019}, which is precisely the gap the LLM strategy selector in Combination B is designed to address.
 
\subsubsection{Unity ML-Agents in Practice}
 
Unity ML-Agents with PPO has been shown to converge to effective seeking behaviour within approximately 200,000 training steps in a single-agent hide-and-seek task \citep{meshram2025}, validating ML-Agents as the RL backbone here. That environment, however, involves a single agent with no opponent strategy and no LLM component, leaving multi-agent coordination and runtime adaptation open.
 
\subsubsection{Evaluation Against Scripted Opponents}
 
DQN opponents have been evaluated in the PettingZoo tag environment against four scripted player strategies, comparing reward distributions with Welch's $t$-test \citep{pillai2023}, a methodology that directly informs the evaluation approach here. However, Welch's $t$-test assumes normally distributed outcomes, unlikely for time-to-defeat, which is bounded by a 120-second ceiling that strong squads are expected to cluster near. This project therefore adopts the Mann-Whitney U test \citep{conover1999} instead, which makes no normality assumption. The closest platform precedent compares PPO against FSM and behaviour-tree opponents in a Unity 3D environment, finding that RL-trained NPCs outperform FSM-based opponents on task success, engagement, and realism \citep{supli2025}, though with only a single NPC and no multi-agent coordination or LLM component.
 
\subsection{Multi-Agent Reinforcement Learning}
\label{sec:literature_multi}
 
Training five agents rather than one is not a free choice; it introduces the non-stationarity problem that the multi-agent RL literature exists to solve, and the coordination mechanism adopted here has to be defended against the alternatives this literature offers rather than assumed to be adequate.
 
\subsubsection{Cooperative MARL and Parameter Sharing}
 
When multiple agents share an environment, the single-agent RL formulation breaks down because the environment becomes non-stationary from each agent's perspective. MARL treats the problem as a stochastic game \citep{busoniu2010}. Full inter-agent communication is impractical in a fast-paced environment, which motivates the parameter-sharing approach adopted here: a single policy network updated by the collective experience of all agents has been shown to converge to effective joint policies on homogeneous cooperative tasks, combined with an agent-indication signal, at substantially lower computational cost than independent per-agent policies \citep{terry2021} --- the direct evidence base for the shared-policy design used here.
 
Training against a diverse set of opponent archetypes, rather than a single fixed opponent, has also been shown to produce policies that generalise substantially better \citep{bansal2018}, motivating the step-based bot-type rotation used during training.
 
\subsubsection{Centralized Training with Decentralized Execution}
 
QMIX implements Centralised Training with Decentralised Execution (CTDE) by decomposing a joint action-value function into per-agent components under a monotonicity constraint, letting cooperative learning signals flow between agents during training without runtime communication \citep{rashid2020} --- the same principle behind the shared policy and jointly applied terminal reward used here. Full CTDE with a centralised critic, however, is a heavier architecture than this project's scope allows, and its absence is acknowledged in Section~\ref{sec:future_work} as a direction for future work.
 
\subsubsection{Hierarchical Multi-Agent Approaches}
 
Hierarchical MARL has been applied to multi-aircraft combat, modelling the scenario as a two-team zero-sum stochastic game with role-conditioned policies trained via Soft Actor-Critic \citep{kong2023}, demonstrating effective team coordination in continuous action spaces structurally analogous to the squad-versus-bot setup here. The key difference this project tests is that the high-level role assignment there is itself a trained component, whereas in Combination B it is a pre-trained language model operating without fine-tuning.
 
\subsection{LLM Integration with Reinforcement Learning}
\label{sec:literature_llm_rl}
 
LLMs have been paired with RL systems before, but in two structurally different ways with very different implications for what such a system can and cannot do at runtime, a distinction that matters directly to whether Combination B's design is a coherent idea in the first place.
 
\subsubsection{LLMs as Reward Designers at Training Time}
 
An LLM has been used to automatically design dense reward functions for Minecraft agents, reporting a 15-point task completion improvement over hand-crafted rewards \citep{li2024auto}, later extended with a self-correcting mechanism that raised a car racing task's success rate from 9\% to 74\% \citep{afonso2025self}. Both show LLMs can reason usefully about agent behaviour from a description of the environment --- but in both cases the LLM operates only at training time, and once training ends the resulting agent has no mechanism for responding to a specific opponent during actual play. This project tests what changes when the LLM instead remains active throughout evaluation, a design neither study explores.
 
\subsubsection{LLMs as Runtime Strategic Advisors}
 
Real-time strategy environments are established in the literature as posing distinctive AI challenges around decision timing and reactivity \citep{ontanon2013}, which is part of the reason the LLM sidecar here is restricted to a coarse five-second polling interval rather than low-level action selection: local LLM inference cannot approach the reactivity a per-frame controller would need, so the two systems are separated by timescale rather than combined into one.
 
GPT-4 has been shown to reach near-Nash-equilibrium play in repeated two-player games and adapt to opponent behaviour across rounds \citep{akata2025playing} --- a necessary precondition for Combination B, since without genuine strategic reasoning there would be no basis for expecting Mistral 7B's tag selections to outperform a random baseline. The effect is clearest for large models, and whether it holds at 7B scale is untested and treated as an open question here. The architectural design of Combination B is most directly informed by a system combining an LLM with an RL policy for the Werewolf social deduction game, where the LLM reasons about hidden information and proposes candidate actions while an RL policy trained via self-play selects the final action \citep{xu2023werewolf} --- the closest existing precedent for pairing runtime LLM reasoning with RL execution, though in a turn-based, text-only domain rather than a real-time spatial one.
 
\subsubsection{Hierarchical LLM-RL Architectures}
 
An LLM-powered agent (Voyager) has been shown to autonomously build a skill library in Minecraft through iterative prompting with no parameter fine-tuning \citep{wang2023voyager}, supporting the feasibility of treating the LLM sidecar here as a black-box component queried through structured prompts. This separation has been formalised directly using a slow LLM agent to decompose tasks and a fast agent to execute the RL pipeline, improving sample efficiency on the MineDojo benchmark \citep{liu2024rlgpt} --- the five-second polling interval used in Combination B is a direct instantiation of this slow-planner, fast-executor separation. Verbal reinforcement has been shown to maintain structured decision output across sequential reasoning steps through prompt design alone \citep{shinn2023reflexion}, supporting the approach adopted here, where Mistral 7B is constrained to a single-digit output $\{0,1,2,3\}$ via the \texttt{num\_predict} Ollama parameter rather than through fine-tuning.
 
\subsection{NPC Behavior Adaptation}
\label{sec:literature_npc}
 
Adaptiveness is easy to want and hard to define. The literature below supplies both the empirical reason to care about it and the specific mechanism --- conditioning a shared policy on an external signal --- that makes one network capable of more than one behaviour without retraining it several times over.
 
\subsubsection{Limitations of Scripted AI}
 
Finite state machines and behaviour trees define fixed transitions between predefined states, letting a player who has encountered a given NPC type reliably anticipate and exploit it. Adaptive opponents that respond to a player's actual performance level have been shown to produce statistically significant improvements in engagement and learning outcomes over static-difficulty conditions \citep{bontchev2024}, motivating the choice of time-to-defeat as the primary metric here.
 
\subsubsection{Conditioning Shared Policies on External Signals}
 
A transformer-based policy receiving a loot-configuration encoding as an additional observation has been shown to generalise across distributional shifts in game parameters without full retraining, producing distinctly different tactical behaviour in response to the conditioning signal \citep{sestini2021}. This principle maps directly onto the strategy-tagging mechanism in Combination B: the conditioning signal must be present and rewarded during training for the policy to use it at evaluation time, which is why Combination B samples the tag randomly during training rather than holding it at zero.
 
\subsubsection{RL-Trained NPCs Versus FSM-Based NPCs}
 
RL-trained NPCs using PPO have independently been reported to achieve an 82\% task success rate against 61\% for FSM-based agents, with higher engagement and realism scores \citep{ayoub2025} --- reaching the same conclusion as the comparison in Section~\ref{sec:literature_rl} \citep{supli2025}. Two independent studies converging on similar numbers gives reasonable confidence that the RL-over-FSM result is robust rather than an artefact of one environment.
 
\subsection{Research Gap}
\label{sec:literature_related}
 
The papers reviewed above establish, with reasonable confidence: RL-trained NPCs outperform FSM-based NPCs \citep{supli2025,ayoub2025}; parameter sharing and CTDE both provide effective coordination mechanisms for cooperative multi-agent tasks, and training against diverse opponents improves policy robustness \citep{busoniu2010,rashid2020,terry2021,bansal2018}; LLMs possess genuine strategic reasoning in abstract or turn-based settings, most clearly at large model scale \citep{akata2025playing,bakhtin2022diplomacy}; LLMs can shape RL training effectively as reward designers \citep{li2024auto,afonso2025self}; a slow LLM planner paired with a fast RL executor is a validated hierarchical pattern \citep{wang2023voyager,liu2024rlgpt,shinn2023reflexion}; and policies conditioned on an external signal can exhibit qualitatively different behaviour without separate training \citep{sestini2021}.
 
Each of these threads, however, has a limitation directly relevant to this project. The MARL and LLM-reward-design literature is convincing on method but silent on runtime adaptiveness, since none of it keeps the LLM active during evaluation \citep{li2024auto,afonso2025self}. The runtime LLM literature is convincing on architecture but confined to turn-based, text-only domains \citep{xu2023werewolf,bakhtin2022diplomacy}. The spatial combat literature is convincing on the value of RL over scripted AI but leaves the policy fixed once training ends \citep{vinyals2019,berner2019,supli2025,ayoub2025}. No single study combines a spatial real-time environment, a runtime LLM strategy layer, and a multi-agent squad.

The specific untested combination is: a pre-trained multi-NPC shared PPO policy, conditioned on runtime strategy tags selected by a locally hosted LLM observing the live game state, evaluated against multiple opponent archetypes in a real-time spatial combat environment. The experimental design in this project isolates the runtime LLM as the sole independent variable, holding every other aspect of the system constant between Combination A and Combination B, so that any observed difference in time-to-defeat can be attributed specifically to the LLM's strategic tag selection.
 
\section{Research Design and Implementation}
\label{sec:method}
 
Almost every component described below exists in its current form because a different approach was tried first and ruled out during development, not because it was the obvious first choice. This section accounts for those decisions and the evidence behind each one.
 
\subsection{System Architecture}
\label{sec:method_overall}
 
Combination A and Combination B differ in exactly one respect, and getting that isolation right so that any measured difference in Section~\ref{sec:results} can be attributed to LLM guidance and nothing else is the central architectural constraint around which this section is organised.
 
\subsubsection{Experimental Design Overview}
 
Two experimental configurations are implemented and compared within a Unity 6.3.3 LTS combat environment.
 
\textbf{Combination A} serves as the RL-only baseline. Five NPC agents share a single PPO policy trained without any LLM involvement. The strategy tag in each agent's 23-float observation vector is held constant at zero throughout training and evaluation.
 
\textbf{Combination B} adds an LLM strategy selector. In training, strategy tags $\{0,1,2,3\}$ are randomly sampled per episode to teach four tactical modes via strategy-conditioned reward shaping. In evaluation, a Python sidecar queries a local Mistral 7B every five seconds, broadcasting the chosen tag via JSON to all five NPC observation vectors in Unity.
 
\begin{figure}[H]
    \centering
    \includegraphics[width=0.98\columnwidth]{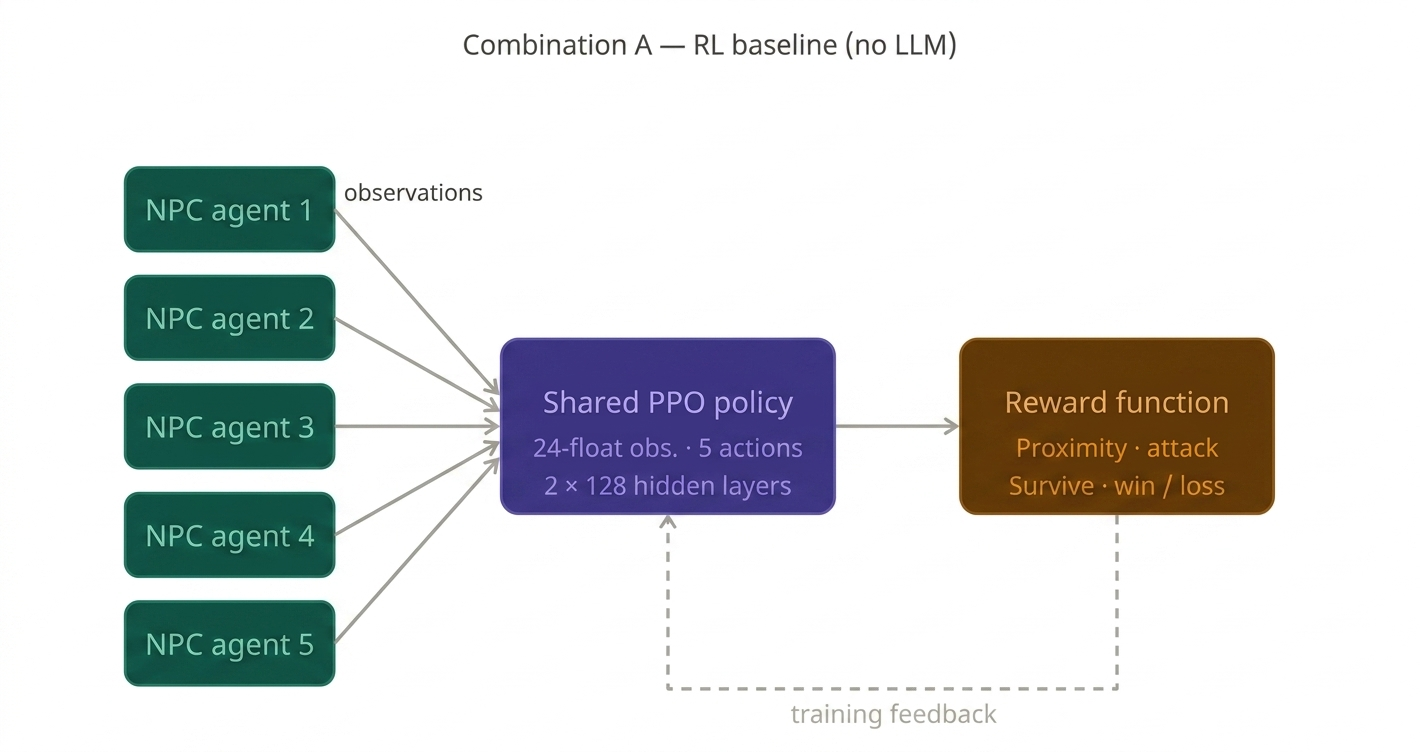}
    \caption{Combination A: RL-only shared policy. Five NPCs share one PPO policy; the strategy tag is fixed at zero.}
    \label{fig:combo_a}
\end{figure}
 
\begin{figure}[H]
    \centering
    \includegraphics[width=0.95\columnwidth]{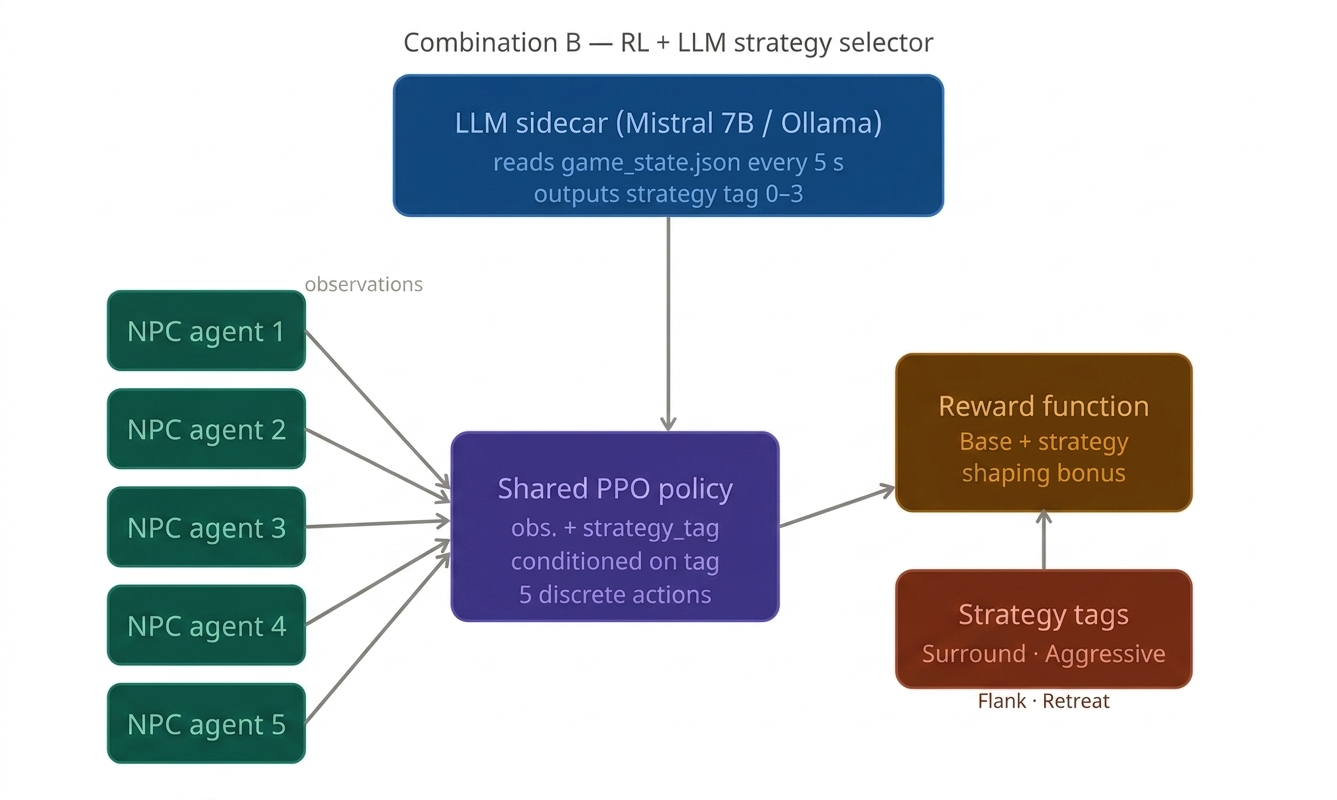}
    \caption{Combination B: RL with LLM strategy selector. The Python sidecar queries Mistral 7B every five seconds and writes a strategy tag conditioning all five NPC observations.}
    \label{fig:combo_b}
\end{figure}
 
The two configurations are structurally identical in all other respects, which includes the same arena, same observation space, same action space, same base reward function, same network architecture, and identical PPO hyperparameters—ensuring that any observed difference in the primary metric can be attributed specifically to the presence or absence of LLM guidance.
 
\subsubsection{Independent and Dependent Variables}
 
The \textbf{independent variable} is the presence of LLM-driven strategy selection, taking two values: absent (Combination A) and present (Combination B). The \textbf{dependent variable} is time-to-defeat, defined as the elapsed simulation time in seconds from episode start to the elimination of all five NPC agents, the elimination of the bot, or the expiry of the 120-second maximum episode duration. Longer survival is operationalized as greater NPC adaptiveness.
 
\subsubsection{Confounding Variable Controls}
 
Arena layout, entity statistics, and PPO hyperparameters are held constant across both combinations and all bot types. Spawn positions are randomized within fixed zones at episode start. Opponent behavior is deterministic within each bot type, eliminating opponent-learning confounds. Reproducibility is ensured by version-controlled YAML configurations and episode-level CSV logging.
 
\subsection{Game Environment Design}
\label{sec:method_design}
 
Every number that follows was chosen to shape what the policy could learn, not for realism's sake—an arena, obstacle layout, or bot HP pool that made the fight too short or too lopsided would have produced a policy with nothing meaningful left to compare in Section~\ref{sec:results}.
 
\subsubsection{Arena Layout}
 
The environment is a flat, enclosed $30 \times 30$ unit plane bounded by four visible BoxCollider walls, implemented in Unity 6.3.3 LTS using the Universal Render Pipeline. Six asymmetrically placed obstacle cubes ($1.5 \times 1.5 \times 1.5$ units) provide potential cover positions and contribute a cover-proximity observation to each NPC's 23-float observation vector. NPCs spawn within a defined zone on the negative-X side; the bot spawns in a separate zone on the positive-X side. Walls were made visible from invisible to aid debugging; this has no effect on agent behavior, which is governed by physics colliders.
 
\begin{figure*}[t]
    \centering
    \includegraphics[width=\textwidth]{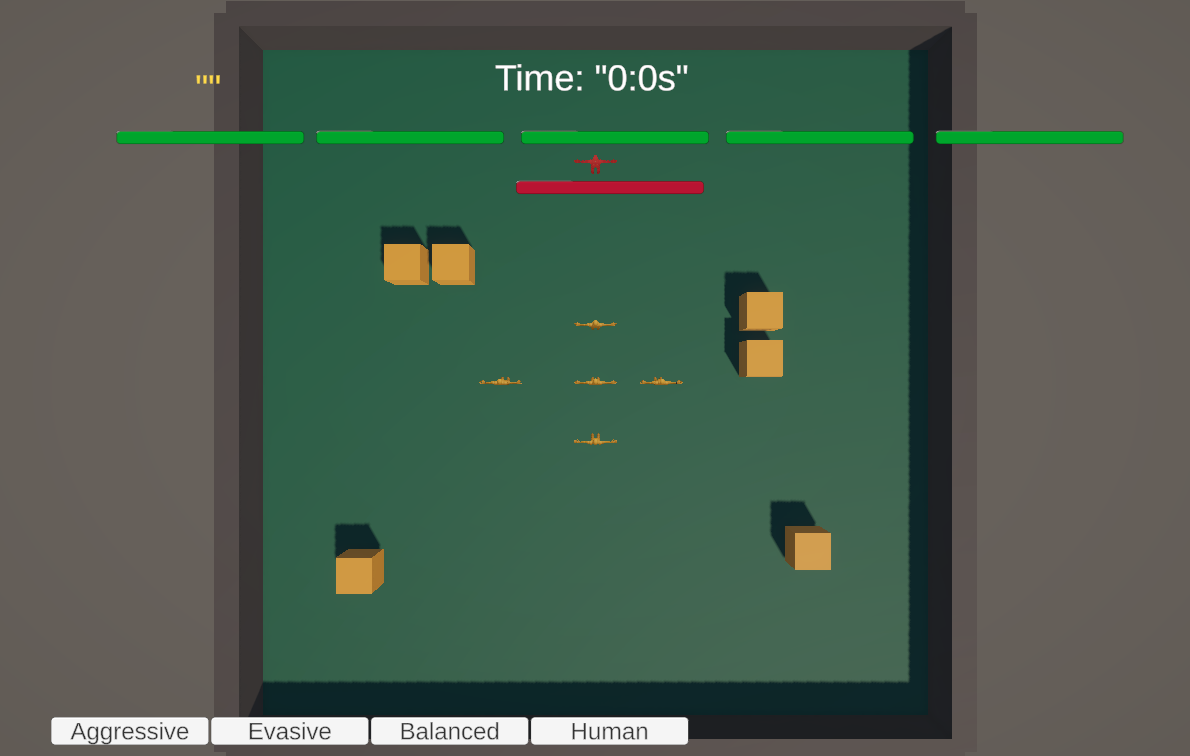}
    \caption{Top-down layout of the $30 \times 30$ combat arena, showing obstacle placement and the NPC/bot spawn zones.}
    \label{fig:arena_layout}
\end{figure*}
 
\noindent The optional human-controlled mode was included solely for personal evaluation and debugging and was not used in the reported experiments or final results.
 
\subsubsection{Entity Statistics}
 
Table~\ref{tab:entity_stats} presents the combat parameters governing the environment. NPC attack damage was increased from 10 to 15 HP and cooldown reduced from 1.0 to 0.5 seconds relative to the initial design specification to bring NPC combat effectiveness closer to parity with the bot and ensure a three-outcome distribution (NPC win, bot win, timeout).
 
\begin{table}[H]
\centering
\caption{Entity combat statistics}
\label{tab:entity_stats}
\small
\renewcommand{\arraystretch}{1.2}
\begin{tabularx}{\columnwidth}{lXXXX}
\toprule
\textbf{Entity} & \textbf{HP} & \textbf{Speed} & \textbf{Range} & \textbf{Dmg / CD} \\
\midrule
Bot   & 300 & 5 u/s & 3.0 u & 15 / 0.8s \\
NPC ($\times5$) & 60  & 4 u/s & 2.5 u & 15 / 0.5s \\
\bottomrule
\end{tabularx}
\end{table}
 
The bot's HP was increased from the 100~HP specified in the initial design to 300~HP because a 100~HP bot would be eliminated in approximately two seconds under maximum simultaneous NPC attack, an episode too short for agents to encounter diverse game states. At 300~HP, episodes last long enough to provide the training-signal diversity needed for policy convergence.
 
\subsubsection{Hardware Specification}
 
Table~\ref{tab:hardware_software} presents the hardware and software configuration used for all training and evaluation.
 
\begin{table}[H]
\centering
\caption{Hardware and software specification}
\label{tab:hardware_software}
\small
\renewcommand{\arraystretch}{1.2}
\begin{tabularx}{\columnwidth}{p{2.6cm}X}
\toprule
\textbf{Component} & \textbf{Specification} \\
\midrule
CPU & AMD Ryzen 7 5000 Series \\
RAM & 16 GB \\
GPU & NVIDIA GeForce RTX 4060 Laptop GPU \\
OS & Windows 11 Home 24H2 \\
Storage & 512 GB SSD + 1 TB HDD \\
Unity Editor & 6.3.3 LTS (URP) \\
Python & 3.10.10 (venv) \\
ML-Agents (Python) & mlagents 1.1.0 \\
ML-Agents (Unity) & 4.0.3 \\
Ollama & 0.30.10 \\
Mistral & mistral:latest via Ollama \\
\bottomrule
\end{tabularx}
\end{table}
 
\subsection{Animation and Character Design}
\label{sec:method_animation}
 
An RL policy only ever outputs a discrete action index; turning that into a character that visibly swings a weapon, and making sure a hit only registers when the animation actually connects, was a separate implementation problem that the reward function has no way of solving on its own.
 
\subsubsection{Character Models and Shared Animator Controller}
 
Characters and animations are sourced from two Unity Asset Store packages by Kevin Iglesias: Human Basic Motions FREE and Human Melee Animations FREE, both sharing a single humanoid avatar definition that makes animations retarget-compatible across all characters without additional configuration. A single Animator Controller is used for both NPC agents and the opponent bot; the difference between the two lies exclusively in what drives the Animator parameters at runtime. Locomotion states are implemented as a 2D Freeform Cartesian blend tree parameterised by SpeedX and SpeedZ, required because the animation set includes directional variants that require two independent parameters to blend correctly.
 
\subsubsection{Animation Event Damage System}
 
Damage is never applied at the moment an attack action is selected in \texttt{OnAction\-Received}. Instead, a Unity Animation Event is embedded at the visual impact frame of each attack clip, invoking \texttt{Trigger\-Damage()} on the CombatHandler component. This ensures damage can only land if the target remains within range at the precise frame of visual contact, maintaining mechanical consistency with the animation.
 
\subsection{Policy Architecture}
\label{sec:policy_architecture}
 
Five separately trained NPCs would have been the more obviously "correct" multi-agent setup; the justification for not doing that, and for what is done instead, is the subject of this section.
 
\subsubsection{Parameter Sharing Across Five NPCs}
 
All five NPC agents share a single PPO policy network through Unity ML-Agents parameter sharing. Each agent has an independent BehaviorParameters component configured with the behaviour name \texttt{NPC\_SharedPolicy}, causing all five agents' transitions to contribute to one set of policy weights per training update. Empirical evidence \citep{terry2021} confirms that parameter sharing consistently outperforms independent policies on cooperative homogeneous multi-agent tasks, directly justifying this design choice over five separate networks. This implementation does not constitute formal MARL: there is no centralised critic, no joint value function, and no explicit inter-agent communication. The mechanism for implicit coordination is the shared terminal rewards for squad win and squad loss.
 
\subsubsection{Network Architecture}
 
The shared policy network consists of three fully connected hidden layers of 256 neurones each with ReLU activations, a softmax output head over six discrete actions, and a separate value network of identical architecture used by the PPO critic. Network capacity was increased from an earlier 128-unit, 2-layer configuration to accommodate the 23-float observation vector's 12-float ally relational component, which requires greater representational capacity to learn coordination-relevant spatial patterns.
 
\subsection{Observation Space}
\label{sec:observation_space}
 
Each NPC agent collects a 23-float observation vector at each decision step, structured across four blocks as detailed in Table~\ref{tab:obs_space}.
 
\begin{table}[H]
\centering
\caption{NPC observation vector (23 floats)}
\label{tab:obs_space}
\small
\renewcommand{\arraystretch}{1.2}
\begin{tabularx}{\columnwidth}{p{1.7cm}p{0.6cm}X}
\toprule
\textbf{Block} & \textbf{Floats} & \textbf{Contents} \\
\midrule
Self         & 3  & Normalised X position, normalised Z position, HP ratio \\
Bot          & 7  & Bot X, bot Z, bot HP ratio, distance to bot (normalised), can-attack binary, nearest cover X, nearest cover Z \\
Allies       & 12 & Four allies $\times$ (X, Z, HP ratio) \\
Strategy tag & 1  & Current tag / 3.0, bounded in $[0,1]$; always 0.0 in Combination A \\
\bottomrule
\end{tabularx}
\end{table}
 
In Combination A the strategy tag is always 0.0; in Combination B training it is sampled uniformly at random per episode and is updated by the LLM sidecar every five seconds during evaluation.
 
\subsection{Action Space}
\label{sec:action_space}
 
The action space is a single discrete branch with six actions, as shown in Table~\ref{tab:action_space}. The ranged action (index 5) was added after initial analysis showed that NPCs with a melee-only action space could not deliver effective damage against the evasive bot, which maintained distance too consistently for melee range to be reached.
 
\begin{table}[H]
\centering
\caption{Discrete action space (branch size = 6)}
\label{tab:action_space}
\small
\renewcommand{\arraystretch}{1.2}
\begin{tabularx}{\columnwidth}{p{0.6cm}p{2.0cm}X}
\toprule
\textbf{Idx} & \textbf{Name} & \textbf{Description} \\
\midrule
0 & Move Forward  & Move toward the bot's current position. \\
1 & Strafe Left   & Move along the local left vector. \\
2 & Strafe Right  & Move along the local right vector. \\
3 & Melee Attack  & Execute melee attack via CombatHandler. \\
4 & Retreat       & Move directly away from the bot. \\
5 & Ranged Attack & Fire orb projectile at the bot via OrbLauncher. \\
\bottomrule
\end{tabularx}
\end{table}
 
\subsection{Reward Function}
\label{sec:reward_function}
 
A proximity reward incentivises sustained engagement:
\begin{equation}
r_{\text{prox}} = 0.01 \times \left(1 - \frac{d_t}{d_{\text{max}}}\right)
\end{equation}
where $d_t$ is the current Euclidean distance to the bot and $d_{\text{max}} = 42.43$ units. A per-step survival bonus $r_{\text{survive}} = +0.001$ prevents suicidal behaviour. Terminal rewards are $r_{\text{attack}} = +0.5$ on confirmed damage, $r_{\text{death}} = -1.0$ on own death, $r_{\text{win}} = +5.0$ to all surviving NPCs when the bot reaches zero HP, and $r_{\text{loss}} = -2.0$ to all NPCs on squad elimination.
 
When strategy shaping is enabled (Combination B), additional per-tag shaping terms are applied in \texttt{OnActionReceived}. Table~\ref{tab:strategy_tags} presents the formulations using halved coefficients adopted after training instability was diagnosed in an earlier run (Section~\ref{sec:combination_b}).
 
\begin{table*}[t]
\centering
\caption{Strategy tags, tactical intent, and shaping formulations (\texttt{comboB\_v02})}
\label{tab:strategy_tags}
\small
\renewcommand{\arraystretch}{1.25}
\begin{tabularx}{\textwidth}{p{0.7cm}p{2.0cm}p{5.5cm}X}
\toprule
\textbf{Tag} & \textbf{Name} & \textbf{Tactical Intent} & \textbf{Shaping Term} \\
\midrule
0 & Surround   & Encircle bot from multiple angles & $r = +0.005 \times |\sin(\theta)|$ \\
1 & Aggressive & Rush and apply sustained pressure & $r = +0.01 \times (1 - d/d_{\text{max}})$ \\
2 & Flank      & Approach from bot's lateral arcs  & $r = +0.0075$ if $60^\circ < |\theta| < 120^\circ$ \\
3 & Retreat    & Pull back and preserve HP         & $r = \min(0.005 \times d/d_{\text{max}},\ 0.005)$; $-0.005$ if $d < 8$ \\
\bottomrule
\end{tabularx}
\end{table*}
 
Here $\theta$ is the signed angle between the bot's forward direction and the vector from the bot to the NPC. The original coefficients (0.01 / 0.02 / 0.015) were halved because the retreat term at its original magnitude could fully cancel the base proximity reward, creating degenerate episodes in which both the NPC squad and the evasive bot retreated simultaneously with no combat engagement. The aggressive coefficient (tag~1) was retained at 0.01, as it was not implicated in the instability.
 
\subsection{Training Configuration}
\label{sec:training_config}
 
The same YAML configuration governs both training runs; all Combination B-specific behaviour is implemented at the script layer rather than in the YAML file.
 
\begin{lstlisting}[language=XML, caption={ML-Agents PPO training configuration (ComboA.yaml and ComboB.yaml are identical)}, label={lst:yaml}]
behaviors:
  NPC_SharedPolicy:
    trainer_type: ppo
    hyperparameters:
      batch_size: 2048
      buffer_size: 81920
      learning_rate: 3.0e-4
      beta: 5.0e-3
      epsilon: 0.2
      lambd: 0.95
      num_epoch: 2
      learning_rate_schedule: linear
    network_settings:
      normalize: true
      hidden_units: 256
      num_layers: 3
    reward_signals:
      extrinsic:
        gamma: 0.99
        strength: 1.0
    max_steps: 3000000
    time_horizon: 128
    summary_freq: 10000
    keep_checkpoints: 5
\end{lstlisting}
 
Both combinations use identical PPO hyperparameters: batch size 2048, buffer size 81920 ($40\times$ batch, ensuring diverse experience before each update in a five-agent environment with a wide joint state space), learning rate $3.0\times10^{-4}$ (linear schedule), clipping $\epsilon=0.2$ following the original PPO specification~\citep{schulman2017}, GAE $\lambda=0.95$, discount $\gamma=0.99$, three hidden layers of 256 units, time horizon 128 (capturing delayed reward consequences inherent in episodic combat), trained for 3,000,000 steps. Observation normalisation is enabled as a safeguard against gradient instability.
 
To ensure the shared policy learns to handle all three opponent profiles, a step-based automatic rotation was implemented using \texttt{Academy.Instance.StepCount}, with each bot type receiving exactly 1,000,000 of the 3,000,000 total steps, following \citep{bansal2018} on the benefits of training against diverse opponent behaviours. The schedule runs aggressive (steps 0--999,999), evasive (steps 1,000,000--1,999,999), and balanced (steps 2,000,000--2,999,999).
 
\subsection{Combination B --- LLM Strategy Selector}
\label{sec:combination_b}
 
One of the two subsections below exists specifically because the first version of this system did not work as intended --- the training instability it describes is as much a part of the design as the parts that worked the first time.
 
\subsubsection{LLM Sidecar Architecture}
 
Communication between Unity and the LLM uses a file-based JSON bridge. Every five seconds, a \texttt{StrategyBridge} component writes the game state to a file and reads back a strategy tag to update all NPCs. As Mistral 7B on CPU typically requires 5--30 seconds per response, a 5-second polling interval is used. Invalid or missing responses retain the previous strategy tag and are logged.
 
\begin{lstlisting}[language=bash, caption={Game state JSON written by Unity every 5 seconds}]
{
  "player_hp": 67.5,
  "avg_npc_hp": 44.2,
  "npcs_alive": 4,
  "bot_moving": true,
  "nearest_dist": 6.3
}
\end{lstlisting}
 
\subsubsection{Prompt Design and Constrained Output}
 
The sidecar converts the game state JSON into a natural language prompt requesting exactly one digit from $\{0,1,2,3\}$, using a constrained \texttt{num\_predict} parameter to keep inference fast. This constrained generation approach follows \citep{shinn2023reflexion}, who show that prompt design alone can produce structured, parseable outputs across sequential decisions without fine-tuning. Mistral 7B was selected because it runs locally via Ollama, ensuring no game state is transmitted to external services.
 
\subsubsection{Training Configuration for Combination B}
 
During training, the strategy tag is sampled uniformly from $\{0,1,2,3\}$ at the start of each episode. The Retreat tag is excluded when the active bot is Evasive or Balanced, following a reward collapse observed at approximately 1,600,000 steps in an earlier training run, caused by both the NPC squad and the Evasive bot retreating simultaneously, generating shaping rewards without combat and corrupting policy learning. This exclusion affects only tag sampling and leaves the PPO hyperparameters, reward function, and observation and action spaces unchanged, preserving the validity of the A-versus-B comparison.
 
\subsection{Opponent Bots}
\label{sec:opponent_bots}
 
Three scripted opponent bots replace a human player to enable controlled, repeatable evaluation. The \textbf{Aggressive Bot} charges the nearest living NPC at full speed, representing a forward-pressure-only opponent. The \textbf{Evasive Bot} continuously moves away from the nearest NPC, attacking only when an NPC closes to melee range, representing a kiting opponent whose challenge is specifically spatial coordination. The \textbf{Balanced Bot} uses three state-dependent modes --- Dominant, Retreat, and Hunting --- switching between them based on its own HP and the number of nearby NPCs, serving as the primary evaluation benchmark because its behaviour changes dynamically during each episode.
 
\subsection{Evaluation Protocol}
\label{sec:evaluation_protocol}
 
Each combination is evaluated against each of the three bot types across 100 episodes, giving 300 episodes per combination and 600 total. Policy weights are frozen during evaluation, and the exported ONNX model runs in inference-only mode. Episodes exceeding 120 seconds terminate with outcome TIMEOUT. The primary metric is time-to-defeat, measured as the elapsed simulation time from episode start to end. Statistical comparison uses the two-sided Mann-Whitney U test with rank-biserial effect size:
\begin{equation}
r = 1 - \frac{2U}{n_1 n_2}
\end{equation}
where $|r|<0.3$ is interpreted as small, $0.3 \leq |r| < 0.5$ as medium, and $|r| \geq 0.5$ as large \citep{conover1999}.
 
\section{Evaluation and Results}
\label{sec:results}
 
Combination A comprises 300 episodes using the \texttt{comboA\_v01} model with the strategy tag fixed at zero, while Combination B comprises 300 episodes using \texttt{comboB\_v02} with the Mistral 7B sidecar querying every five seconds.
 
\subsection{Training Convergence Summary}
\label{sec:training_summary}
 
Combination A trained cleanly over 3,000,000 steps in approximately 5.25 hours, reaching a smoothed cumulative reward of 56.04 with stable policy loss between 0.016 and 0.019 throughout. Episode length smoothed at 1,086 steps (approximately 22 seconds simulation time), confirming that the policy learned meaningful combat engagement behaviour.
 
Combination B required two training runs. The initial run experienced reward collapse at approximately 1.6 million steps due to degenerate dynamics when the Retreat tag was sampled against the Evasive bot. The corrected run (\texttt{comboB\_v02}) completed 3,000,000 steps in approximately 6.64 hours, reaching a smoothed reward of 44.74. Value loss ended at 0.19 versus 0.08 for Combination A, reflecting the added complexity of learning four tag-conditioned behaviours. No further reward collapse occurred.
 
\subsection{Descriptive Statistics}
\label{sec:descriptive_stats}
 
Table~\ref{tab:comboa_desc} presents the full descriptive statistics for Combination A across the three opponent bot types, and Table~\ref{tab:combob_desc} presents the corresponding statistics for Combination B.
 
\begin{table*}[t]
\centering
\caption{Combination A --- descriptive statistics ($n=100$ per bot type)}
\label{tab:comboa_desc}
\small
\renewcommand{\arraystretch}{1.2}
\begin{tabularx}{\textwidth}{lXXXXXX}
\toprule
\textbf{Bot Type} & \textbf{Mean (s)} & \textbf{Median (s)} & \textbf{Std (s)} & \textbf{NPC Win \%} & \textbf{Bot Win \%} & \textbf{Timeout \%} \\
\midrule
Aggressive & 77.72 & 80.03 & 22.51 & 8.0\% & 90.0\% & 2.0\% \\
Evasive    & 78.31 & 73.42 & 21.12 & 90.0\% & 1.0\% & 9.0\% \\
Balanced   & 104.08 & 120.00 & 22.21 & 11.0\% & 29.0\% & 60.0\% \\
\bottomrule
\end{tabularx}
\end{table*}
 
\begin{table*}[t]
\centering
\caption{Combination B --- descriptive statistics ($n=100$ per bot type)}
\label{tab:combob_desc}
\small
\renewcommand{\arraystretch}{1.2}
\begin{tabularx}{\textwidth}{lXXXXXX}
\toprule
\textbf{Bot Type} & \textbf{Mean (s)} & \textbf{Median (s)} & \textbf{Std (s)} & \textbf{NPC Win \%} & \textbf{Bot Win \%} & \textbf{Timeout \%} \\
\midrule
Aggressive & 66.54 & 61.98 & 26.51 & 8.0\% & 84.0\% & 8.0\% \\
Evasive    & 70.48 & 67.15 & 16.21 & 93.0\% & 3.0\% & 4.0\% \\
Balanced   & 105.82 & 120.01 & 22.71 & 24.0\% & 14.0\% & 62.0\% \\
\bottomrule
\end{tabularx}
\end{table*}
 
\begin{figure}[H]
    \centering
    \includegraphics[width=0.95\columnwidth]{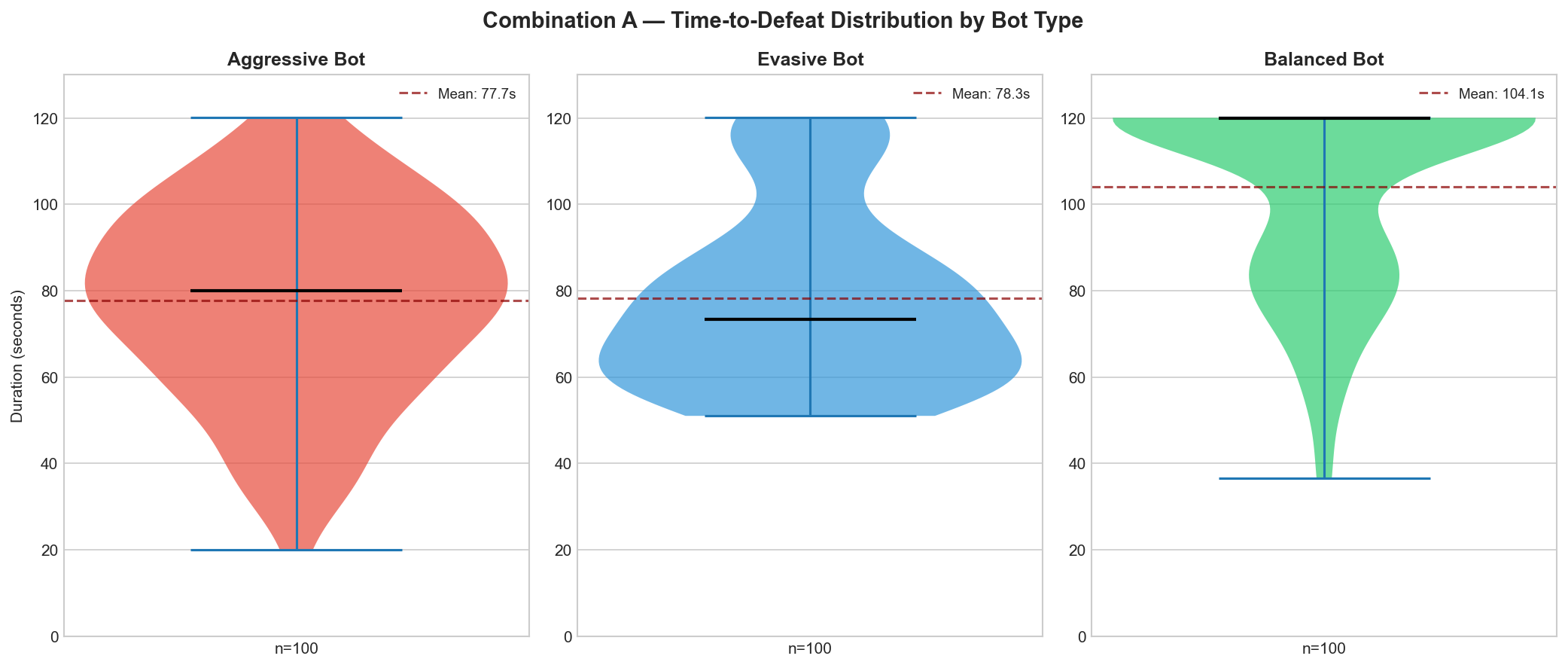}
    \caption{Combination A time-to-defeat distributions by bot type (dashed: mean; solid: median).}
    \label{fig:violin_comboa}
\end{figure}

Figure~\ref{fig:violin_comboa} shows the time-to-defeat distributions for Combination A across the three bot types. The Aggressive condition produces a roughly symmetric distribution centred around 80 seconds, reflecting the bot's consistent forward-pressure behaviour. The Evasive condition produces a bimodal distribution with a large cluster of NPC squad wins between 55--80 seconds and a smaller cluster of timeouts at 120 seconds. The Balanced condition is dominated by the 120-second ceiling, with 60\% of episodes reaching timeout and the remaining outcomes spread across the full duration range.
 
\begin{figure}[H]
    \centering
    \includegraphics[width=0.95\columnwidth]{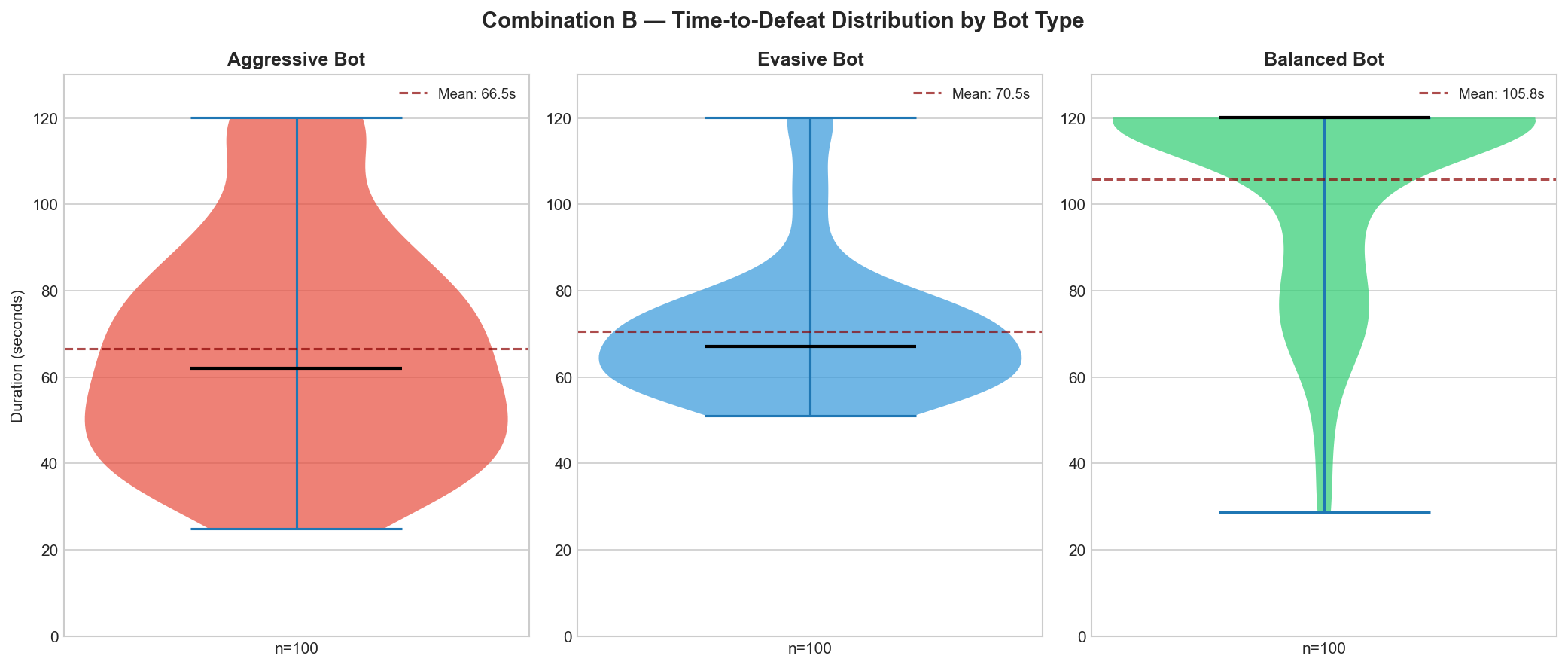}
    \caption{Combination B time-to-defeat distributions by bot type (dashed: mean; solid: median).}
    \label{fig:violin_combob}
\end{figure}

Figure~\ref{fig:violin_combob} shows the corresponding distributions for Combination B. The Aggressive condition shows a wider, flatter distribution with a lower mean than Combination A, reflecting more short losses alongside the increased timeout rate (8\% vs 2\%). The Evasive condition is notably tighter and left-shifted relative to Combination A, reflecting that NPC squad wins are being achieved more quickly (mean 67.1 seconds in Combination B versus 73.8 seconds in Combination A). The Balanced condition closely resembles Combination A, with a slight rightward shift in mean (105.82 versus 104.08 seconds).
 
The most notable differences are: the Evasive NPC win rate increases from 90\% to 93\%; the Aggressive bot win rate decreases from 90\% to 84\% with a corresponding increase in timeouts from 2\% to 8\%; and the Balanced NPC win rate more than doubles from 11\% to 24\%, with a corresponding reduction in bot win rate from 29\% to 14\%.
 
\begin{figure}[H]
    \centering
    \includegraphics[width=0.95\columnwidth]{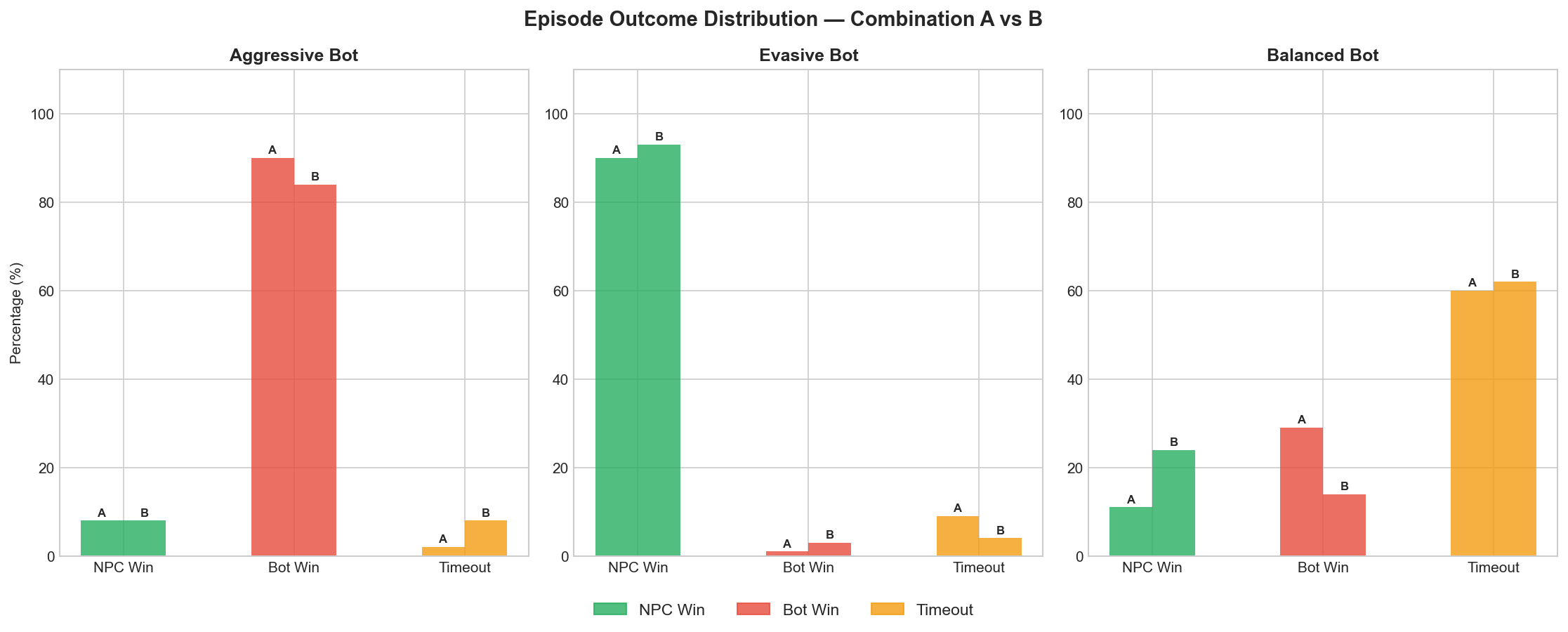}
    \caption{Episode outcome distribution (NPC win, bot win, timeout) by combination and bot type.}
    \label{fig:outcomes_bar}
\end{figure}
 
\subsection{Statistical Analysis}
\label{sec:stats}
 
Table~\ref{tab:mannwhitney} presents the Mann-Whitney U test results for each bot-type condition.
 
\begin{table}[H]
\centering
\caption{Mann-Whitney U test, A vs B (two-sided, $\alpha=0.05$, $n_1=n_2=100$)}
\label{tab:mannwhitney}
\small
\renewcommand{\arraystretch}{1.25}
\begin{tabularx}{\columnwidth}{lXXXX}
\toprule
\textbf{Bot} & \textbf{U} & \textbf{p} & \textbf{r} & \textbf{Direction} \\
\midrule
Aggressive & 6430.5 & 0.0005*** & $-0.286$ & B shorter \\
Evasive    & 6032.0 & 0.0117*   & $-0.206$ & B shorter \\
Balanced   & 3668.5 & 0.0011**  & $+0.266$ & B longer \\
\bottomrule
\end{tabularx}
\end{table}
 
\begin{figure}[H]
    \centering
    \includegraphics[width=0.95\columnwidth]{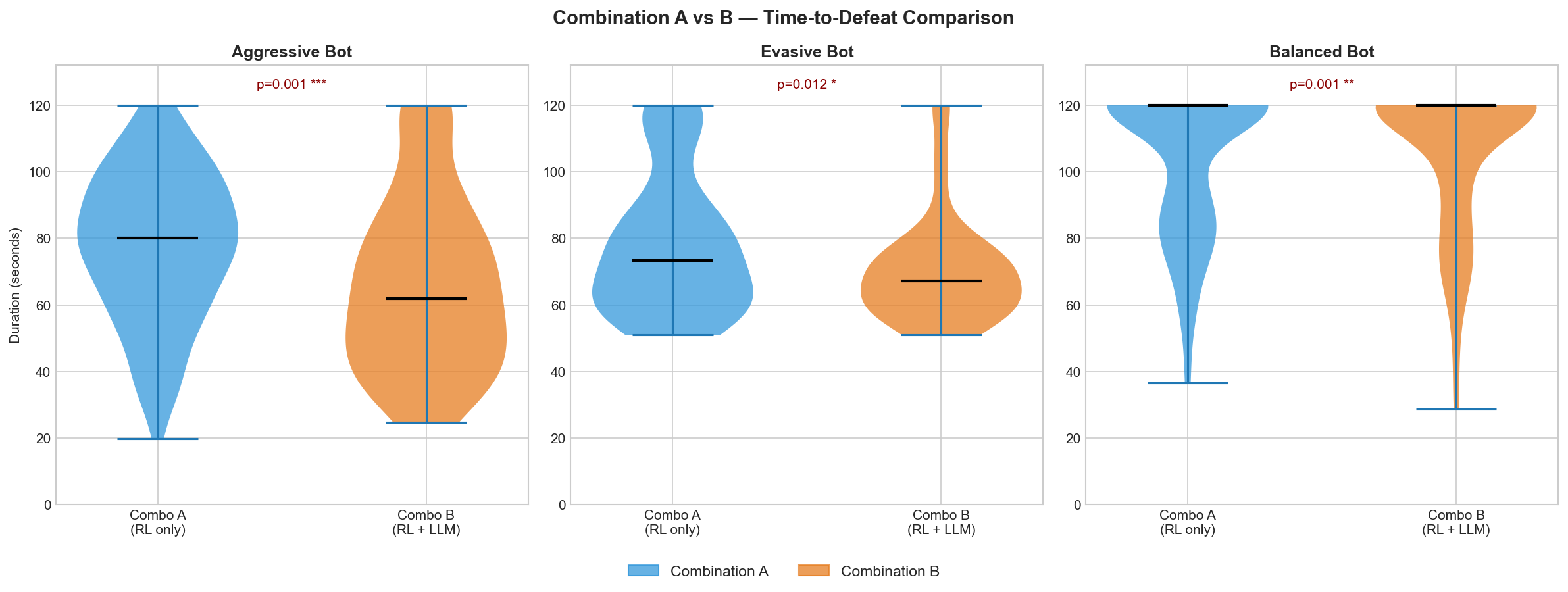}
    \caption{Combination A vs B time-to-defeat, with Mann-Whitney $p$-values annotated per bot type.}
    \label{fig:violin_comparison}
\end{figure}
 
\subsection{LLM Strategy Tag Analysis}
\label{sec:tag_analysis}
 
The strategy decision log recorded 2,430 Mistral 7B strategy selections across the 300 Combination B evaluation episodes (Table~\ref{tab:tag_dist}).
 
\begin{table}[H]
\centering
\caption{LLM strategy tag selection frequency (2,430 total decisions)}
\label{tab:tag_dist}
\small
\renewcommand{\arraystretch}{1.2}
\begin{tabularx}{\columnwidth}{lXXX}
\toprule
\textbf{Tag} & \textbf{Name} & \textbf{Count} & \textbf{\%} \\
\midrule
0 & Surround    & 2036 & 83.8 \\
1 & Aggressive  & 103  & 4.2 \\
2 & Flank       & 107  & 4.4 \\
3 & Retreat     & 184  & 7.6 \\
\bottomrule
\end{tabularx}
\end{table}
 
\begin{figure}[H]
    \centering
    \includegraphics[width=0.95\columnwidth]{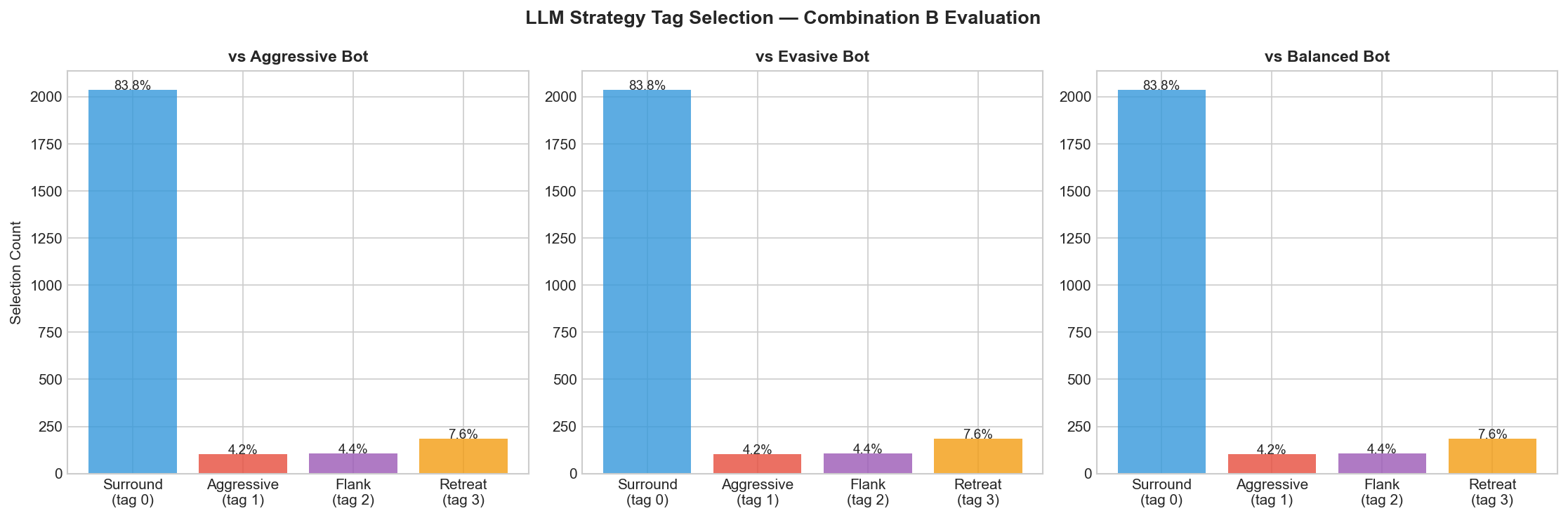}
    \caption{LLM strategy tag selection frequency during Combination B evaluation, by bot type.}
    \label{fig:tag_dist}
\end{figure}
 
The tag distribution reveals an important finding: Mistral 7B selected the Surround tag in 83.8\% of all decisions, with this proportion remaining approximately constant across all three opponent bot types. The 'Retreat' tag was selected in 7.6\% of decisions, exclusively in contexts where mean NPC HP was critically low (mean 2.1\%), consistent with the tag's tactical intent but representing a small fraction of the total. A strategy selector that is effectively reading opponent state and differentiating its response would be expected to produce meaningfully different tag distributions across conditions. The observed uniformity suggests that Mistral 7B, at 7B parameter scale and with the prompt design used here, defaulted to a near-constant 'Surround' preference rather than dynamically adapting to opponent behaviour.
 
\subsection{Interpretation and Verdict}
\label{sec:results_verdict}
 
The three Mann-Whitney U tests each reached statistical significance ($p<0.05$), meaning H0 is rejected in all three conditions. The direction of the effect, however, differs across conditions.
 
Against the \textbf{Aggressive} bot, H0 is rejected ($p=0.0005$, $r=-0.286$) but in the direction \emph{against} H1: Combination B produced shorter mean time-to-defeat (66.54s vs 77.72s) with identical NPC win rates (8\% each). The bot win rate decreased from 90\% to 84\%, but the timeout rate increased from 2\% to 8\%, suggesting the LLM's persistent Surround tag caused NPCs to spread laterally rather than engage the advancing bot directly.
 
Against the \textbf{Evasive} bot, H0 is rejected ($p=0.0117$, $r=-0.206$), again in the direction against the strict H1 definition: episodes were shorter on average (70.48s vs 78.31s). This result requires careful interpretation, however: the NPC win rate increased from 90\% to 93\%, and when episodes ended in NPC wins the mean kill time was faster in Combination B (67.1s vs 73.8s), with more NPCs surviving at the point of kill (4.94 vs 4.10). The shorter time-to-defeat is driven by faster, cleaner kills, not by the squad being eliminated sooner.
 
Against the \textbf{Balanced} bot, H0 is rejected ($p=0.0011$, $r=+0.266$) and in this condition the direction \emph{does} favour H1: Combination B produced longer mean time-to-defeat (105.82s vs 104.08s), and the NPC win rate more than doubled from 11\% to 24\%, with bot win rate falling from 29\% to 14\%.
 
The strict H1 as formulated is therefore partially supported: one of three conditions produces a statistically significant improvement in time-to-defeat in the expected direction.
 
\subsection{Critical Evaluation}
\label{sec:critical_eval}
 
The most significant finding beyond the primary hypothesis test is the LLM tag distribution: Mistral 7B selected the Surround tag in 83.8\% of all decisions, with little variation across bot types, suggesting limited zero-shot generalisation for strategic reasoning. This behaviour explains the differing results across bot types: Surround is effective against a retreating opponent such as the Evasive bot, naturally supporting encirclement, but is less effective against the Aggressive bot, where NPCs spread laterally instead of concentrating their attack.
 
A methodological limitation is that the 83.8\% Surround dominance means many Combination B episodes were effectively running a near-constant Surround tag rather than dynamic, state-responsive tagging. This makes it impossible to fully attribute the observed differences to LLM \emph{reasoning} as opposed to a near-static Surround tag --- a much simpler intervention. Future work should examine whether a hardcoded Surround tag produces similar results to isolate the language model's specific contribution.
 
During training the strategy tag was sampled uniformly at random per episode, but during evaluation the LLM selected tag 0 in 83.8\% of decisions. This creates a training-evaluation distribution mismatch: whether the policy learned distinct tag-0 behaviour versus a general average behaviour cannot be determined from the evaluation data alone.
 
Four further methodological limitations qualify the interpretation of the results. First, the sample size of 100 episodes per condition was not determined by formal power analysis. Second, the primary metric of time-to-defeat does not capture episode quality from a player-experience perspective. Third, the five-second polling interval constrains how rapidly the LLM can respond to within-episode state changes. Fourth, all results were obtained from a single trained model under a single LLM configuration.
 
\section{Discussion}
\label{sec:discussion}
 
The condition-dependent verdict above is only useful if the mechanism behind it is understood well enough to act on. Surround's angular-spread reward happens to counter kiting (the Evasive bot) and partially offsets single-target focus fire (the Balanced bot's Hunting mode), but against an opponent that simply charges the nearest NPC (the Aggressive bot), spreading out is precisely the wrong response, trading concentrated damage for worse individual exposure. The Retreat tag's HP-triggered use against the Balanced bot is the one clear instance of the LLM responding to state rather than defaulting. What this shows is not that the LLM reasoned correctly three times and incorrectly once, but that one fixed default was luckily aligned with two opponent types and badly misaligned with a third.
 
Two explanations for the Surround dominance are plausible. First, the prompt described the available tactics but provided only HP, distance, and movement status, without explicitly identifying the opponent's behavioural mode, so the model may have defaulted to a generally safe strategy. Second, Mistral 7B may lack the strategic reasoning capacity required for this task, consistent with \citep{akata2025playing}, who reported improved performance with larger models. Both explanations indicate opportunities for improving the system design rather than fundamental limitations of the LLM-as-strategy-selector paradigm.
 
The experimental result neither validates nor refutes the paradigm in general, since it characterises it under specific conditions. The paradigm is most likely to succeed when the LLM's default strategic preference happens to be appropriate for the opponent type, as was the case here for the Evasive and Balanced bots, and will underperform when the LLM's default preference is mismatched to the opponent, as with the Aggressive bot. This suggests effective deployment requires either a larger, more capable model that can genuinely differentiate strategies by opponent behaviour, or a prompt design that provides more explicit opponent characterisation to break the Surround dominance. Neither change requires altering the RL policy or the reward function --- they are prompt engineering changes, one of the architectural advantages of the approach.
 
\subsection{Limitations and Future Work}
\label{sec:future_work}
 
Several concrete extensions follow directly from the limitations above. A \textbf{hardcoded Surround baseline} is the most immediate experiment: if it matches Combination B's results, the LLM's contribution was coincidental rather than reasoning-based. \textbf{Richer opponent characterisation in the prompt} --- adding explicit behaviour labels such as whether the bot is advancing, retreating, or switching modes --- would provide Mistral with the information necessary to differentiate tag selections by opponent type. A \textbf{larger or fine-tuned LLM} would test whether Surround dominance is a property of 7B-scale models or of the prompt design. A \textbf{formal MARL implementation} with a centralised critic (e.g. QMIX \citep{rashid2020} or COMA \citep{foerster2018}) would provide explicit coordination signals during training and could be compared against the parameter-sharing baseline to quantify the coordination deficit the LLM tags were partially compensating for. \textbf{Mid-episode tag resampling during training} would align the training distribution with the five-second evaluation protocol and may produce more responsive tag-conditioned behaviour. Finally, a \textbf{player perception evaluation} would connect the statistical results to the player-experience motivation stated in Section~\ref{sec:intro}, which time-to-defeat alone cannot capture.
 
\section{Conclusion}
\label{sec:conclusion}
 
This paper investigated whether a locally hosted large language model can improve NPC adaptiveness by acting as a runtime strategy selector without modifying reinforcement learning policy weights. The proposed system combines a shared PPO policy controlling five NPC agents with a Mistral 7B sidecar that queries the game state every five seconds and broadcasts a strategy tag to all agents.
 
Evaluation across 600 episodes, two system configurations, and three opponent bot types produced statistically significant differences in every condition ($p<0.05$). Against the Balanced bot, the LLM-augmented system achieved longer time-to-defeat and more than doubled the NPC win rate, supporting H1. Against the Evasive bot, it produced faster eliminations and a higher win rate but shorter episode durations, indicating improved tactical efficiency rather than increased survival. Against the Aggressive bot, the LLM's persistent selection of the Surround strategy reduced performance, resulting in shorter episodes than the RL-only baseline.
 
Strategy tag analysis showed that Mistral 7B selected the Surround tag in 83.8\% of decisions with little variation across opponent types, indicating limited zero-shot adaptation to different opponent behaviours. These results demonstrate that LLM-guided runtime strategy selection is feasible and produces measurable behavioural effects, but its effectiveness depends on the model's ability to reason about the current strategic context. Improving prompt design, enriching opponent-state information, or adopting larger models are promising directions for future work. The diagnosis of the initial training instability and its corrected implementation also provide a practical contribution on strategy-conditioned reward shaping in multi-opponent reinforcement learning, complementing existing reward shaping literature \citep{ng1999potential,devlin2012}.
 
\nocite{hu2024survey,liu2024llmagents,ye2025,zheng2025mcu,juliani2018,ontanon2013,schulman2017,sutton1998,vaswani2017attention,bcs2022code}
 
\balance
\bibliographystyle{ACM-Reference-Format}
\bibliography{references}
 
\section*{Bibliography}
\begingroup
\renewcommand{\refname}{}

\endgroup
 
\appendix
 
\section{Agent Prompt Details}
\label{app:prompt}
 
Table~\ref{tab:prompt_details} documents the full prompt structure passed to Mistral 7B at each five-second polling interval.
 
\begin{table}[H]
\centering
\caption{Combination B LLM sidecar prompt details}
\label{tab:prompt_details}
\small
\renewcommand{\arraystretch}{1.3}
\begin{tabularx}{\columnwidth}{p{2.2cm}X}
\toprule
\textbf{Component} & \textbf{Content} \\
\midrule
Role & You control a squad of NPC soldiers fighting an enemy bot. Choose the best squad tactic for the current situation. \\
State & Enemy bot HP, average NPC HP, NPCs alive, bot moving (bool), nearest NPC distance. \\
Action & Reply with exactly one digit: 0 = Surround, 1 = Aggressive, 2 = Flank, 3 = Retreat. \\
Generation & \texttt{num\_predict: 3}, \texttt{temperature: 0.1}, \texttt{top\_p: 0.9}, \texttt{timeout: 40s}. First character parsed; invalid output retains previous tag. \\
Fallback & Empty, non-numeric, or out-of-range output retains the previous strategy tag and is logged. \\
\bottomrule
\end{tabularx}
\end{table}
 
\section{Reproducibility Statement}
\label{app:reproducibility}
 
The full project repository, including Unity scene files, C\# agent and bot scripts, the Python LLM sidecar, YAML training configurations, and the statistical analysis script, is available at: \url{https://github.com/HRITHIKA-NAIR/llm-rl-npc-f21mp}.
 
To reproduce the evaluation results from trained models: install Unity 6.3.3 LTS and open the project; install Python 3.10.10 and the required dependencies; install Ollama and pull \texttt{mistral}; load the evaluation scene with \texttt{trainingMode=false} and select the target bot type; for Combination B, run the LLM sidecar script alongside the Unity scene; run 100 episodes per bot type, with metrics logged automatically to CSV; run the analysis script to reproduce all statistical results and figures.
 
To reproduce training from scratch, run \texttt{mlagents-learn} with the Combination A and Combination B YAML configurations respectively. Note that the Combination B run requires the Retreat-tag exclusion logic to be active against the Evasive bot.
 
\end{document}